\documentclass[letterpaper, 10 pt, conference]{ieeeconf}  

\IEEEoverridecommandlockouts                              
\usepackage[utf8]{inputenc}
\usepackage[T1]{fontenc}
\usepackage{verbatim}
\usepackage{bm}
\usepackage{amsmath}
\usepackage{graphicx}
\usepackage{caption2}
\usepackage{subfigure}
\usepackage{algorithm}
\usepackage{algorithmic}
\usepackage{multirow}
\usepackage{hyperref}
\usepackage{float}
\usepackage{flushend}
\usepackage{amssymb}
\usepackage{makecell}
\usepackage{booktabs}

\title{\LARGE \bf
Interpretable Stochastic Model Predictive Control using Distributional Reinforced Estimation for Quadrotor Tracking Systems
}

\author{Yanran Wang, James O'Keeffe, Qiuchen Qian and David Boyle
}

\begin{document}

\maketitle
\thispagestyle{empty}
\pagestyle{empty}

\begin{abstract}

This paper presents a novel trajectory tracker for autonomous quadrotor navigation in dynamic and complex environments. The proposed framework integrates a distributional Reinforcement Learning (RL) estimator for unknown aerodynamic effects into a Stochastic Model Predictive Controller (SMPC) for trajectory tracking. Aerodynamic effects derived from drag forces and moment variations are difficult to model directly and accurately. Most current quadrotor tracking systems therefore treat them as simple `disturbances' in conventional control approaches. We propose Quantile-approximation-based Distributional Reinforced-disturbance-estimator, an aerodynamic disturbance estimator, to accurately identify disturbances, i.e., uncertainties between the true and estimated values of aerodynamic effects. Simplified Affine Disturbance Feedback is employed for control parameterization to guarantee convexity, which we then integrate with a SMPC to achieve sufficient and non-conservative control signals. We demonstrate our system to improve the cumulative tracking errors by at least 66\% with unknown and diverse aerodynamic forces compared with recent state-of-the-art. Concerning traditional Reinforcement Learning's non-interpretability, we provide convergence and stability guarantees of Distributional RL and SMPC, respectively, with non-zero mean disturbances.
\end{abstract}

\section{INTRODUCTION}
Accurate trajectory tracking for autonomous Unmanned Aerial Vehicles (UAVs), such as quadrotors, is necessary for maintaining autonomy. Although industrial applications of autonomous UAVs, such as commercial deliveries, search-and-rescue \cite{mishra2020drone} and wireless power transfer \cite{qian2022practical}, have attracted much attention in recent years, precisely tracking high-speed and high-acceleration UAV trajectories is an extremely challenging control problem, particularly in unknown and dynamic environments with unpredictable aerodynamic forces.

To achieve safe, precise and reliable quadrotor trajectory tracking, there are two main problems that need to be solved: How can we achieve robust and feasible estimation (or modelling) of the aerodynamic effects on quadrotors in complex dynamic environments? And; How can the whole control framework be integrated with aerodynamic effect estimation to solve the uncertainties and disturbances while tracking trajectory references precisely and reliably?

Previous work has shown that the primary source of uncertainties are aerodynamic effects deriving from drag forces and moment variations caused by the rotors and the fuselage \cite{torrente2021data}. Prominent aerodynamic effects appear at flight speeds of 5 $ms^{-1}$ in wind tunnel experiments by \cite{faessler2017differential}. These effects acting on quadrotors are chaotic and hard to model directly, as they are generated from a combination of the individual propellers and airframe \cite{hoffmann2007quadrotor}, turbulent effects caused by rotor–rotor and airframe–rotor interactions \cite{russell2016wind}, and the propagation of other turbulence \cite{kaya2014aerodynamic}.

Most current approaches to quadrotor trajectory tracking treat aerodynamic effects as simple external disturbances, and do not account for higher-order effects or attempt to deviate from a determined plan \cite{tal2020accurate,ru2017nonlinear,bicego2020nonlinear}. While these solutions are efficient and feasible for lightweight on-board computers, aggressive maneuvers at high speed, e.g., greater than 5 $ms^{-1}$, introduce large positional and attitude tracking errors. Recent data-driven approaches, such as Gaussian Processes (GP) \cite{torrente2021data,wang2022kinojgm} and neural networks \cite{spielberg2021neural} combined with Model Predictive Control (MPC), show accurate modelling of aerodynamic effects. However, due to the nonparametric nature of GP, the GP-based approaches perform poorly in complex environments - where large datasets contain drastic changes in wind speed and heading. In these instances, learning-based (neural networks) approaches perform better than those GP-based approaches \cite{loquercio2021learning}. Achieving adaptability and robustness in complex environments is still challenging, however, primarily because training datasets are collected from simulated platforms and real-world historical records that do not fully describe the complex environments.

In comparison to existing data-driven approaches, Reinforcement Learning (RL), an interactive learning process, is able to learn complex and changeable disturbances - i.e., the errors between the true and estimated values - using much less model information \cite{zhang2021model}. The key challenge of most existing RL approaches \cite{lillicrap2015continuous} is that policy optimization biases toward actions with high variance value estimates, since some of these values will be overestimated by random chance \cite{ma2021conservative}. In risk-sensitive or safety-critical applications such as autonomous quadrotor navigation these actions should be avoided. Recent work on distributional RL \cite{bellemare2017distributional} was proposed to approximate and parameterize the entire distribution of future rewards, instead of the expected value. Distributional RL algorithms have been operated to achieve advanced results on continuous control domains \cite{hessel2018rainbow}. In princple, they provide more complete and richer value-distribution information to enable a more stable learning process \cite{bellemare2017distributional}. Previous distributional RL algorithms parameterize the policy value distribution in different ways, including canonical return atoms \cite{bellemare2017distributional}, the expectiles \cite{rowland2019statistics}, the moments \cite{nguyen2021distributional}, and the quantiles \cite{dabney2018distributional,zhang2019quota}. The quantile approach is especially suitable for autonomous UAV trajectory tracking due to its risk-sensitive policy optimization.

Robust MPC for tracking control of uncertain systems like quadrotors is rapidly developing thanks to advances in hardware and algorithmic efficiency \cite{nguyen2021model}. The robust control approach enables a `worst case' formulation to analyze the stability and performance of a system under bounded uncertainties and disturbances \cite{zhou1998essentials}. However, with consideration of the uncertainties in real-world scenarios, such worst-case design renders the optimal control actions inherently inadequate and overly conservative in practise \cite{mayne2016robust,mesbah2016stochastic}. To avoid the conservatism of the worst case design, Stochastic MPC (SMPC) \cite{schwarm1999chance} uses the probabilistic descriptions, such as stochastic constraints (also called chance constraints), to predict probability distributions of system states within acceptable levels of risk in the receding-horizon optimization \cite{primbs2009stochastic}.

The core challenges for SMPC include: 1) optimizing the feedback control laws over arbitrary nonlinear functions \cite{munoz2020convergence}; 2) the chance constraints are non-convex and intractable \cite{mesbah2016stochastic,zhang2021stochastic}; and 3) the computational complexity will grow dramatically as more uncertainties are added. To address the first challenge, one solution is to use affine parameterization of the control policy over finite horizons. However, this approach cannot guarantee convexity, i.e., the second challenge, where the policy set may still be convex \cite{lofberg2003approximations}. Another solution is an Affine Disturbance Feedback (ADF) control parameterization, proposed in \cite{goulart2008input}. This ADF control parameterization can address the first two challenges, which are optimizing the dynamic function and guaranteeing the decision variables to be convex, respectively. However, the main weakness is that the computational complexity grows quadratically with the prediction horizon, i.e., the third challenge. To overcome this difficulty, a Simplified Affine Disturbance Feedback (SADF) proposed in \cite{zhang2021stochastic}, where the SADF is equivalent to ADF but a finite-horizon optimization can be computed more efficiently using CasADi \cite{andersson2019casadi}, a nonlinear MPC solver. \cite{zhang2021stochastic} achieve good results by implementing SADF with zero-mean disturbance, however it is unclear how the SADF would perform on systems with non-zero-mean disturbances, such as a quadrotor.

To address the two stated issues, we propose \textbf{Qua}ntile-approximation-based \textbf{D}istributional \textbf{Re}inforced-\textbf{D}isturbance-estimation for Stochastic MPC (QuaDRED-SMPC), a systematic, safe and feasible quadrotor trajectory tracking framework for use with high variance aerodynamic effects. The details are as follows:
\begin{enumerate}
    \item Aerodynamic Disturbance Estimator: a Quantile-approximation-based Distributional Reinforced-disturbance-estimator (QuaDRED), described by Algorithm~\autoref{DRL_QuaDRED}, is proposed for aerodynamic disturbance estimation.
    QuaDRED builds upon prior QR-DQN \cite{dabney2018distributional} and QUOTA \cite{zhang2019quota} insofar as QuaDRED is a quantile-approximated distributional RL which uses a set of quantiles to approximate the full value distribution. In Section \uppercase\expandafter{\romannumeral4}-A, theoretical guarantees on convergence of the QuaDRED are provided based on policy evaluation (\textit{Proposition 3}) and policy improvement (\textit{Proposition 4}), respectively. 
    
    \item Trajectory Tracker: Similar to \cite{zhang2021stochastic}, a Simplified Affine Disturbance Feedback (SADF) is used for control parameterization in SMPC (Algorithm~\autoref{SADF_SMPC}), where the convexity can be guaranteed in this process \cite{lofberg2003approximations} and computational complexity can be reduced. Different from prior work assuming zero mean disturbance, we consider the control performance and stability  under non-zero-mean disturbance. We use an Input-to-State Stability (ISS) \cite{jiang2001input} property to find conditions that imply stability and convergence of the tracker.  
    \item The QuaDRED-SMPC framework is proposed to track quadrotor trajectory accurately under high variance aerodynamic effects. The overall control framework is shown in Fig.~\ref{RL_Control_framework}. In Section \uppercase\expandafter{\romannumeral4}-B, the closed-loop stability of the QuaDRED-SMPC framework is demonstrated under Lipschitz Lyapunov function \cite{rifford2000existence}.
\end{enumerate}

Our contributions can be summarized as follows:
\begin{itemize}
\item[1)]
QuaDRED, a distributional RL with quantile approximation that can sufficiently estimate variable aerodynamic disturbances. In the cases tested, we show that QuaDRED outperforms traditional RL, such as Deep Deterministic Policy Gradient (DDPG) \cite{lillicrap2015continuous}, and prior Distributional RL approaches, such as C51 \cite{bellemare2017distributional}. 
\end{itemize}
\begin{itemize}
\item[2)]
The integration of a trajectory tracker with an aerodynamic disturbance estimator, a quadrotor trajectory tracking framework that integrates QuaDRED into a stochastic optimal control problem.
\end{itemize}
\begin{itemize}
\item[3)]
Convergence and stability guarantees: mathematical proofs are provided for the convergence of distributional-RL-based estimator, and the closed-loop stability of stochastic-MPC-based tracker with consideration of non-zero-mean and bounded disturbances.
\end{itemize}

\section{Problem Formulation}
A quadrotor dynamic model has six Degrees of Freedom (DoF), i.e., three linear motions and three angular motions \cite{torrente2021data}. We consider a nonlinear discrete system of quadrotor dynamics with state $\bm{x}\in\mathbb{X}\subseteq\mathbb{R}^{n}$, an additive disturbance $\bm{w}\in\mathbb{W}\subseteq\mathbb{R}^{n_w}$, and control input $\bm{u}\in\mathbb{R}^{n_u}$, defined for all time steps $k\in\mathbb{N}$ by:
\begin{equation}
\bm{x}_{k+1} = \bm{f}(\bm{x}_{k},\bm{u}_{k}, {\bm{e_{f}}}_k) + \bm{w}_k
\label{quadrotor_nonlinear_discrete_system}
\end{equation}
where $\bm{x}_{k}$ and $\bm{u}_{k}$ are the discrete-time state and input vectors of the quadrotor dynamic model. $\bm{w}_{k}$ is a disturbance caused by aerodynamic effects. $\bm{f}(\bm{x},\bm{u}, \bm{e}_{f})$ in \autoref{quadrotor_dynamics} is the continuous-time nominal model of the quadrotor integrating the aerodynamic effect $\bm{e}_{f}$. The state and input vectors of the nominal model are $\bm{x} = [\bm{P}_{WB},\bm{V}_{WB},\bm{q}_{WB},\bm{\omega} _{B}]^T$ and $\bm{u} = T_i, \forall i\in(0,3)$. $\bm {P}_{WB}$, $\bm{V}_{WB}$ and $\bm{q}_{WB}$ are the position, linear velocity and orientation of the quadrotor, and $\bm{\omega}_{B}$ is the angular velocity \cite{wang2022kinojgm}.
\begin{equation}
\begin{aligned}
&\dot{\bm{P}}_{WB} = \bm{V}_{WB}\\
&\dot{\bm{V}}_{WB} = \bm{g}_W + \frac{1}{m}(\bm{q}_{WB}\odot \bm{c} + \bm{e}_f)\\
&\dot{\bm{q}}_{WB} = \frac{1}{2}\Lambda(\bm{\omega} _{B})\bm{q}_{WB}\\
&\dot{\bm{\omega}} _{B} = \bm{J}^{-1}(\bm{\tau}_B-\bm{\omega}\times J\bm{\omega} _{B})
\end{aligned}
\label{quadrotor_dynamics}
\end{equation}
where $\bm{c}$ is the collective thrust $\bm{c} = [0,0,\sum T_i]^{\rm{T}}$ and $\bm{\tau}_B$ is the body torque; $\bm{g}_W = [0,0,-g]^{\rm{T}}$. The operator $\odot$ denotes a rotation of the vector by the quaternion. The skewsymmetric matrix $\Lambda(\bm{\omega})$ is defined in \cite{wang2022kinojgm}. 

We linearize and reformulate \cite{goulart2008input} \autoref{quadrotor_nonlinear_discrete_system} for MPC over a finite horizon $N$:
\begin{equation}
\bm{x}_{t} = \bm{A}x_{0|t} + \bm{B}\bm{u}_{t} + \bm{G}\bm{w}_{t}
\label{linear_discrete_system}
\end{equation}
where $\bm{x}_{t}=[x_{0|t}^{\rm{T}}, x_{1|t}^{\rm{T}}, ..., x_{N|t}^{\rm{T}}]^{\rm{T}}$ and $\bm{u}_{t}=[u_{0|t}^{\rm{T}}, u_{1|t}^{\rm{T}}, ..., u_{N|t}^{\rm{T}}]^{\rm{T}}$ are the sequential states and inputs, and $\bm{w}_{t}=[w_{0|t}^{\rm{T}}, w_{1|t}^{\rm{T}}, ..., w_{N|t}^{\rm{T}}]^{\rm{T}}$ denotes a sequential stochastic disturbance over a horizon of $N$. $\bm{A}$, $\bm{B}$ and $\bm{G}$ are matrices defined in \cite{zhang2021stochastic}.

The following assumptions are made: 

\textit{Assumption 1}: Matrix G is column full rank.

\textit{Assumption 2}: The aerodynamic effect $\bm{e_{fk}}$ is available with no delay at each sampling timestamp.

\textit{Proposition 1}: There exists a control law $\bm{u}_b$ that ensures the nominal model $\bm{f}(\bm{x}_{k},\bm{u}_{k}, \bm{e_{fk}})$ is ISS if the stochastic disturbance $\bm{w}_t$ is independent and identically distributed (i.i.d.) a zero-mean distribution, i.e., $\mathbb{E}{(w_k)}=0$.

\textit{Proof}: Based on \textit{Assumption 2}, the nominal model $\bm{f}(\bm{x}_{k},\bm{u}_{k}, \bm{e_{fk}})$ is seen as $\bm{f}(\bm{x}_{k},\bm{u}_{k}^{'})$, in which the aerodynamic force $\bm{e_{fk}}$ is a constant term. Then we prove $\bm{f}(\bm{x}_{k},\bm{u}_{k}^{'})$ is ISS \cite{jiang2001input}. According to \cite{munoz2020convergence} and \cite{goulart2008input}, there exists a continuous function $V_b$: $\mathbb{X}\rightarrow\mathbb{R_+}$ that is an ISS-Lyapunov function. Therefore, the nominal model $\bm{f}(\bm{x}_{k},\bm{u}_{k}, \bm{e_{fk}})$ with $\mathbb{E}{(w_k)}=0$ is ISS. \hfill $\blacksquare$

\section{QuaDRED-SMPC}
In this section, we present the proposed QuaDRED-SMPC control framework. Traditional non-interacted methods, e.g., Gaussian Process \cite{torrente2021data} and RDRv \cite{faessler2017differential}, are insufficient for quadrotor dynamic disturbance estimation. This work addresses the limitation, and proposes a novel and feasible disturbance estimation with continuous environmental interactions for variable winds.

\subsection{Quantile-approximation-based Distributional Reinforced-disturbance-estimation \label{D_RL}}

We consider a distributional Bellman equation \cite{bellemare2017distributional}, the aim of which is different from traditional RL, i.e., maximizing the expectation of value-action function $Q$. In the policy evaluation setting, given a deterministic policy $\pi$, the state-action distribution $Z^\pi$ and the \textit{Bellman operator} $\mathcal{T}^\pi$ are defined as \cite{bellemare2017distributional,dabney2018distributional}:
\begin{equation}
\mathcal{T}^{\pi} Z(\bm{s},\bm{a})\overset{D}{:=} R(\bm{s},\bm{a})+\gamma Z(\bm{s'},\bm{a'})
\label{Bellman_Pred_D_RL}
\end{equation}
where $\bm{s}\in S$, $\bm{a}\in A$, $\bm{p}\in P: S\times A\times S$, $R$ and $\bm{\gamma}\in[0,1]$ are the state vector, action vector, transition probability, immediate reward function and discount rate, respectively, in a tuple Markov Decision Process \cite{puterman2014markov}: $MDP:=\left \langle S,A,P,R,\gamma \right \rangle$. $\pi$ is a stationary policy mapping one state $s \in S$ to one action $a \in A$. In the control setting, a distributional \textit{Bellman optimality operator} $\mathcal{T}$ with quantile approximation is also proposed in \cite{dabney2018distributional}:
\begin{equation}
\begin{aligned}
\mathcal{T} Z(\bm{s},\bm{a})\overset{D}{:=} R(\bm{s},\bm{a})&+\gamma Z(\bm{s'},{\rm{arg}} \underset{a'}{max}\underset{\bm{p},R}{\mathbb{E}}[Z(\bm{s'},\bm{a'})])\\
Z_\theta(\bm{s},\bm{a})&:=\frac{1}{N}\sum\limits_{i=1}^{N} \delta_{q_i(\bm{s},\bm{a})}
\label{Bellman_Control_D_RL}
\end{aligned}
\end{equation}
where $Z_\theta \in Z_Q$ is a quantile distribution mapping one state-action pair $(s,a)$ to a uniform probability distribution supported on $q_i$. $Z_Q$ is the space of quantile distribution within $N$ supporting quantiles. $\delta_z$ denotes a Dirac with $z \in \mathbb{R}$. The state-action value $Q$ is then approximated by $Q_{j|K} \overset{D}{:=}\frac{1}{K}\sum\limits_{k=(j-1)K+1}^{(j-1)K+K}q_k(s,a)$. These quantile approximations -i.e., $\left\{q_i\right\}$ - are operated based on Quantile Huber Loss.

The Wasserstein Metric, also known as the Mallows metric, is a true probability metric with no disjoint support issues, and therefore suitable to calculate the metric distance between the target $\mathcal{T}^{\pi} Z$ and the prediction $Z$. A contraction is proved in \cite{dabney2018distributional} over the Wasserstein Metric:
\begin{equation}
\overset{-}{d}_{\infty}(\Pi_{W_1}\mathcal{T}^{\pi}Z_1,\Pi_{W_1}\mathcal{T}^{\pi}Z_2)\leq \overset{-}{d}_{\infty}(Z_1,Z_2)
\label{contraction_WM}
\end{equation}
where $W_p$, $p\in[1,\infty]$ denotes the $p$-Wasserstein distance. $\overset{-}{d}_{p}:={\rm{sup}}W_p(Z_1,Z_2)$ denotes the maximal form of the  $p$-Wasserstein metrics. $\Pi_{W_1}$ is a quantile approximation under the minimal 1-Wasserstein distance $W_1$.

The aim of our proposed QuaDRED is to track the trajectory reference $\bm{x}_{m,t}$ generated from Kino-JSS \cite{wang2022kinojgm} accurately,  therefore the immediate reward $r_{t+1}$ is defined as:
\begin{equation}
r_{t+1}=-(\bm{x}_{t}-\bm{x}_{m,t})^{\rm{T}}H_1 (\bm{x}_{t}-\bm{x}_{m,t})-\bm{u}_{t}^{\rm{T}}H_2\bm{u}_{t}
\label{reward}
\end{equation}
where $H_1$ and $H_2$ are positive definite matrices. Then we use DDPG architecture \cite{lillicrap2015continuous} for the continuous and high-dimensional disturbance estimation.

\subsection{Control Parameterization}
A SADF \cite{zhang2021stochastic}, as shown in \autoref{affine_fb_SADF}, is an equivalent and tractable formulation of the original affine feedback prediction control policy proposed in \cite{jiang2001input}. More importantly, the SADF has fewer decision variables which can decrease computational complexity and improve calculation efficiency.
\begin{equation}
\bm{u}_{i|t}=\sum\limits_{k=0}^{i-1} \bm{M}_{i-k|t}\bm{w}_{k|t}+\bm{v}_{i|t}
\label{affine_fb_SADF}
\end{equation}
where the $\bm{M}_{t}$ is a lower block diagonal Toeplitz structure. $i\in\mathbb{N}_{[1,N-1]}, j\in\mathbb{N}_{i-1}$ and the open-loop control sequence $\bm{v}_{i|t} \in \mathbb{R}, i\in\mathbb{N}_{N-1}$ are decision variables at each time step $t$.

According to \cite{goulart2008input}, the predicted cost can be transformed as:
\begin{equation}
\begin{aligned}
&\mathcal{L}(\bm{x}_t,\bm{u}_t)=\mathcal{L}_N(\bm{x}_t,\bm{M}_t,\bm{v}_t)\\
&=\left \|H_{x}x+H_{u}\bm{v}\right \|^{2}_2+\mathbb{E}[\left \|(H_{u}{\bm{M}}\mathcal{G}+H_{w})\bm{w}\right \|^{2}_2]
\end{aligned}
\label{predicted_cost}
\end{equation}
where $H_{x}$ and $H_{u}$ are coefficient matrices which are constructed from \autoref{predicted_cost}. $\mathcal{G}:=I_N \bigotimes G$ denotes Kronecker product of matrices $I_N$ and $G$. For the convexity guarantee, the matrix $(H_{u}{\bm{M}}\mathcal{G}+H_{w})$ is positive semidefinite.
Thus, the optimal control problem, reformulated by SADF (\autoref{affine_fb_SADF}) is as follows:
\begin{equation}
\begin{aligned}
\underset{\bm{M}_t,\bm{v}_t}{\rm{min}} \mathcal{L}_N(\bm{x}_t,\bm{M}_t&,\bm{v}_t), {\rm{s.t.}} \forall w_{i|t}\in \mathbb{W}, \forall i\in \mathbb{N}_{N-1}\\
\rm{subject}\quad \rm{to}\quad &\bm{x}_{t} = \bm{A}x_{0|t} + \bm{B}\bm{u}_{t} + \bm{G}\bm{w}_{t}\\
&\bm{u}_{i|t}=\sum\limits_{k=0}^{i-1} \bm{M}_{i-k|t}\bm{w}_{k|t}+\bm{v}_{i|t}\\
&H_{u}{\bm{M}}\mathcal{G}+H_{w} \geq 0\\
&(\bm{x}_t,\bm{u}_t)\in \mathbb{Z}\\
&\bm{x}_{N|t}\in \mathbb{X}_{f}\\
&\bm{x}_{0|t}=\bm{x}_t\\
\end{aligned}
\label{SMPC_Formulation}
\end{equation}

The optimal control problem is a strictly convex quadratic program or second-order cone program (SOCP) if $\mathbb{W}$ is a polytope or ellipsoid when $\mathbb{Z}$ and $\mathbb{X}_{f}$ are polytopic \cite{goulart2008input}. In this case, this problem can be seen as deterministic MPC with nonlinear constraints, which can be solved by some nonlinear MPC solvers, e.g., CasADi \cite{andersson2019casadi} and ACADOS \cite{verschueren2018towards}.

\subsection{Quantile-approximation-based Distributional Reinforced-disturbance-estimation for SMPC}
\begin{figure}[]
  \centering
  \includegraphics[scale=0.54]{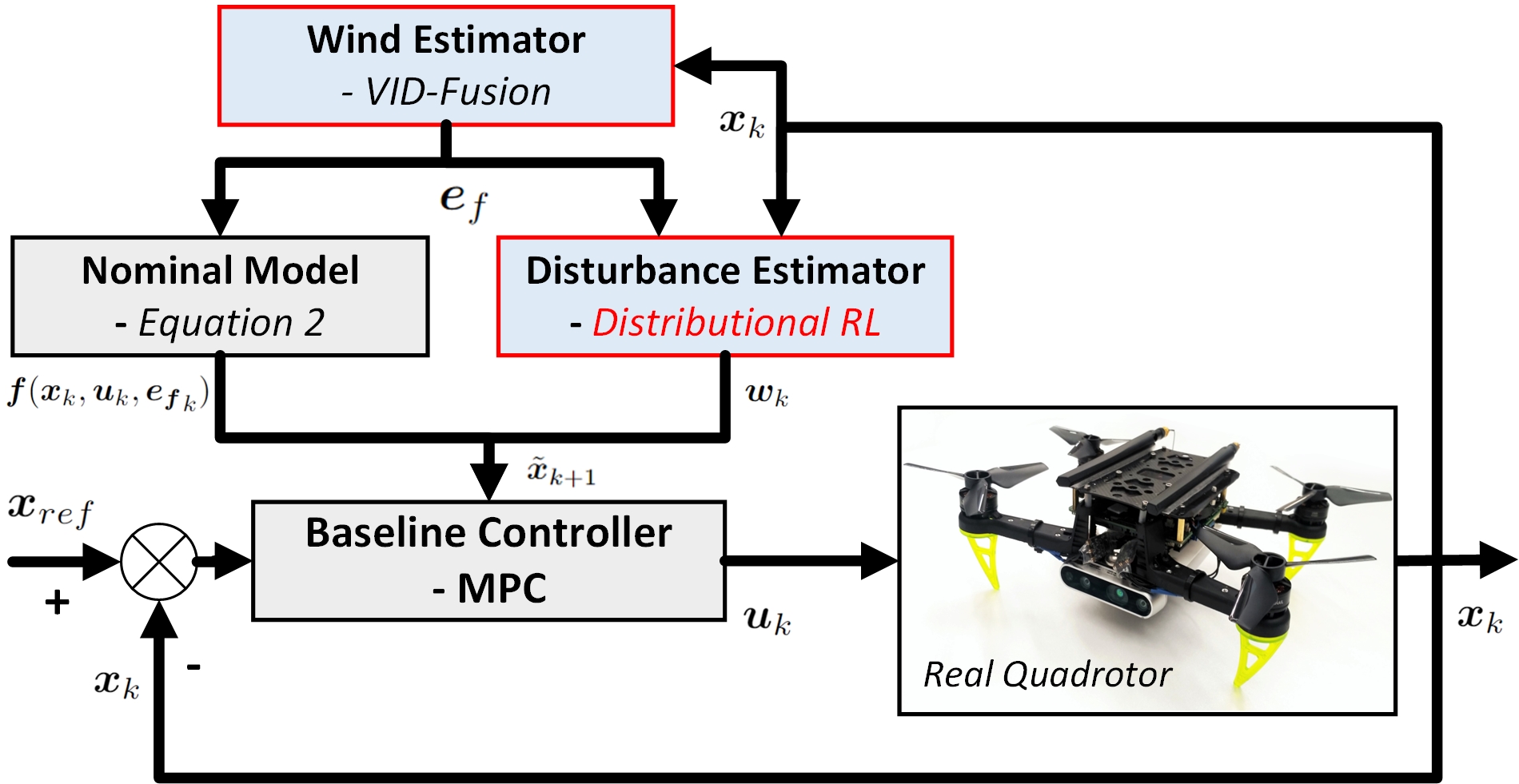}
  \caption{QuaDRED-SMPC}
  \label{RL_Control_framework}
\end{figure}

The objective of this work is to design a quadrotor controller achieving accurate aerodynamic effect estimation, which we define as combined wind estimation and aerodynamic disturbance estimation, for tracking the reference state $\bm{x}_{ref}$ of the nominal model (\autoref{quadrotor_dynamics}). The overall control framework for the quadrotor is shown in Fig.~\ref{RL_Control_framework}. The SADF in stochastic MPC and QuaDRED are shown in Algorithm~{\autoref{SADF_SMPC}} and Algorithm~{\autoref{DRL_QuaDRED}}, respectively.

\begin{algorithm}[t]
\caption{SADF-SMPC}
\begin{algorithmic}[1]
\STATE \textbf{Get}:\\
    - the reference data $\bm{x}_{ref}$ from the quadrotor trajectory planning and generation module, i.e., Kino-JSS \cite{wang2022kinojgm}\\
    - the measurement state $\bm{x}_k$ from on-board sensors\\
    - the wind estimation $\bm{e}_{fk}$ from VID-Fusion \cite{ding2020vid}\\
\STATE \textbf{Initialize}:\\
    - the parameters $\theta^\mu$ and $\theta^Q$ for the actor $\mu$ and the critic $Q$, respectively\\
    - the decision variables $\bm{M}_{0}$ and $\bm{v}_{0}$ in \autoref{affine_fb_SADF}\\
    - the initial state $\bm{s}_{0}$\\
\FOR{each sampling timestamps $k$}
\STATE \textbf{Repeat}
\STATE $\bm{s}_k \gets \left [ \bm{x}_k, \bm{e}_{fk} \right ]$
\STATE Select an action vector $\bm{w}_k \gets [w_{0|k}^{\rm{T}}, w_{1|k}^{\rm{T}}, ..., w_{N|k}^{\rm{T}}]^{\rm{T}}$ from $\bm{w}_k=\bm{a}_k \gets \mu(\bm{a}_k|\bm{s}_k)$ in QuaDRED (Algorithm~\autoref{DRL_QuaDRED})
\STATE $\bm{u}_{i|k} \gets \sum\limits_{l=0}^{i-1} \bm{M}_{i-l|k}\bm{w}_{l|k}+\bm{v}_{i|k}$
\STATE $\bm{u}_{k} \gets \bm{u}_{0|k}$, $\bm{w}_{k} \gets \bm{w}_{0|k}$
\STATE $\bm{x}_{k} \gets \bm{A}x_{0|k} + \bm{B}\bm{u}_{k} + \bm{G}\bm{w}_{k}$
\STATE Solve optimization problem \autoref{SMPC_Formulation} with nonlinear MPC solver\\
\STATE \textbf{Until} convergence
\STATE $\bm{u}_{k} \gets \bm{v}_{0|k}$
\STATE $\bm{x}_{k+1}$, $\bm{e}_{fk+1} \gets {\rm{Real Quadrotor}}(\bm{u}_{k})$
\STATE $\bm{s}_{k+1} \gets [\bm{x}_{k+1}, \bm{e}_{fk+1}]$
\STATE $\bm{x}_{ref} \gets$ Kino-JSS
\STATE $k \gets k+1$
\ENDFOR
\end{algorithmic}
\label{SADF_SMPC}
\end{algorithm}

\begin{algorithm}[]
\caption{QuaDRED}
\hspace*{0.02in} {\bf Input:}
$\bm{s}_k$, $\bm{s}_{k+1}$, $\bm{u}_{k}$, $\theta^\mu$, $\theta^Q$\\
\hspace*{0.02in} {\bf Output:}
$\bm{a}_k$
\begin{algorithmic}[1]
\STATE \textbf{Initialize}:\\
    - $\theta^{\mu^t} \gets \theta^\mu$, $\theta^{Q^t} \gets \theta^Q$ update the target parameters from the predicted parameters\\
    - the replay memory $D \gets D_{k-1}$\\
    - the batch $B$, and its size\\
    - a small threshold $\xi \in \mathbb{R_+}$ \\
    - the random option selection probability $\epsilon$
    - the option termination probability $\beta$ \\
    - quantile estimation functions $\left \{ q_i   \right \}_{i=1,...,N}$\\
\STATE \textbf{Repeat}
\FOR{each sampling step from $D$}
\STATE Select a candidate option $z_k$ from $\left \{z^0, z^1, ..., z^M   \right \}$
\STATE $z_k \gets 
\begin{cases}
z_{k-1} & {w.p.\;1-\beta} \\
{\rm{random \; option}} & {w.p.\;\beta\epsilon} \\
{\rm{arg max}}_{z}Q(\bm{s}_k,z) & {w.p.\;\beta(1-\epsilon)}
\end{cases}$
\STATE Execute $w_k$, get reward $r_k$ and the next state $\bm{s}_{k+1}$
\STATE $D.\bf{Insert}([\bm{s}_{k}, \bm{u}_{k}, r_k, \bm{s}_{k+1}])$
\STATE $B \gets D.\bf{sampling}$
\STATE $y_{k,i} \gets \rho_{\tau_i}^{\mathcal{K}}(r_k+\gamma q_i'(\bm{s}_{k+1},w_k^*)$
\STATE $J_{\theta^\mu} \gets \frac{1}{N}\sum\limits_{i=1}^{N}\sum\limits_{i'=1}^{N}[y_{k,i'}-q_i(\bm{s}_k,w_k)]$
\STATE $y \gets \beta {\rm{arg max}}_{z'}Q(\bm{s}_{k+1},z')+(1-\beta)Q(\bm{s}_{k+1}.z_k)$
\STATE $J_{\theta^Q} \gets (r_t+\gamma y-Q(\bm{s}_t,z_t))^2$
\STATE $\theta^\mu \gets \theta^\mu - l_{\mu}\nabla_{\theta^\mu} J_{\theta^\mu}$
\STATE $\theta^Q \gets \theta^Q - l_{\theta} \nabla_{\theta^Q} J_{\theta^Q}$
\ENDFOR
\STATE \textbf{Until} convergence, i.e., $J_Q^\theta < \xi$
\end{algorithmic}
\label{DRL_QuaDRED}
\end{algorithm}

\section{Properties of QuaDRED-SMPC}
In this section, the properties of the proposed control framework QuaDRED-SMPC are analyzed, including convergence of QuaDRED and stability guarantees of the Controller SADF-SMPC.
\subsection{Convergence Analysis of QuaDRED}
We present the following \textit{Proposition 3} and \textit{Proposition 4} on the convergence analysis for the Distributional RL (QuaDRED) in Section \uppercase\expandafter{\romannumeral4}-A.

\textit{Lemma 2} (\cite{bellemare2017distributional}): The \textit{Bellman operator} $\mathcal{T}^\pi$ is a $p$-contraction under the $p$-Wasserstein metric $\overset{-}{d}_{p}$.

\textit{Lemma 2} suggests that an effective way in practice to minimize the Wasserstein distance between a distribution $Z$ and its Bellman update $\mathcal{T}^{\pi}Z$ can be found in \autoref{Bellman_Pred_D_RL}, which attempts iteratively to minimize the $L2$ distance between $Z$ and $\mathcal{T}^{\pi}Z$ in Temporal Difference learning.

\textit{Proposition 3 (Policy Evaluation)}: Let $\Pi_{W_1}$ be a quantile approximation under the minimal 1-Wasserstein distance $W_1$, $\mathcal{T}^\pi$ be the \textit{Bellman operator} under a deterministic policy $\pi$ and $Z_{k+1}(\bm{s},\bm{a})=\Pi_{W_1}\mathcal{T}^{\pi} Z_k(\bm{s},\bm{a})$. The sequence $Z_{k}(\bm{s},\bm{a})$ converges to a unique fixed point $\overset{\sim}{Z_\pi}$ under the maximal form of $\infty$-Wasserstein metric $\overset{-}{d}_{\infty}$.

\textit{Proof}: 
\autoref{contraction_WM} implies that the combined operator $\Pi_{W_1} \mathcal{T}^\pi$ is an $\infty$-contraction \cite{dabney2018distributional}. We conclude using Banach’s fixed point theorem that $\mathcal{T}^\pi$ has a unique fixed point, i.e., $\overset{\sim}{Z_\pi}$. Furthermore, \autoref{Bellman_Control_D_RL} implies that all moments of $Z$ are bounded. Therefore, we conclude that the sequence $Z_{k}(\bm{s},\bm{a})$ converges to $\overset{\sim}{Z_\pi}$ in $\overset{-}{d}_{\infty}$ for $p\in[1,\infty]$. \hfill $\blacksquare$ 

\textit{Proposition 4 (Policy Improvement)}: Let $\bm{\pi}_{\bm{old}}$ be an old policy, $\bm{\pi}_{\bm{new}}$ be a new policy and $Q(s, a)=\mathbb{E}[Z(s,a)]$ in \autoref{Bellman_Control_D_RL}. There exists $Q^{\bm{\pi}_{\bm{new}}}(s, a) \geq Q^{\bm{\pi}_{\bm{old}}}(s, a)$, $\forall s\in \mathcal{S}$ and $\forall a\in \mathcal{A} $.

\textit{Proof}: Based on \autoref{Bellman_Control_D_RL}, there exists:
\begin{equation}
\begin{aligned}
V^\pi(s_{t})&=\mathbb{E}_\pi {Q^\pi(s_{t},\pi(s_{t}))}\\
&\leq \underset{a\in \mathcal{A}}{\rm{max}} \mathbb{E}_\pi {Q^\pi(s_{t},a)} \\
&=\mathbb{E}_{\pi'}{Q^\pi(s_{t},{\pi'}(s_{t}))}
\end{aligned}
\label{Q_V}
\end{equation}
where $\mathbb{E}_\pi[\cdot]=\sum_{a\in A}\bm{\pi}(a|s)[\cdot]$, and $V^\pi(s)=\mathbb{E}_\pi \mathbb{E}[Z_k(s,a)]$ is the value function. According to \autoref{Q_V} and \autoref{Bellman_Control_D_RL}, it yields:
\begin{equation}
\begin{aligned}
Q^{\bm{\pi}_{\bm{old}}} &=Q^{\bm{\pi}_{\bm{old}}}(s_{t},\bm{\pi}_{\bm{new}}(s_{t})) \\
&= r_{t+1}+\gamma \mathbb{E}_{s_{t+1}} \mathbb{E}_{\bm{\pi}_{\bm{old}}} Q^{\bm{\pi}_{\bm{old}}}(s_{t+1},{\bm{\pi}_{\bm{old}}}(s_{t+1}))\\
&\leq r_{t+1}+\gamma \mathbb{E}_{s_{t+1}} \mathbb{E}_{\bm{\pi}_{\bm{new}}}{Q^{\bm{\pi}_{\bm{old}}}(s_{t+1},{\bm{\pi}_{\bm{new}}}(s_{t+1}))}\\
&\leq r_{t+1}+\mathbb{E}_{s_{t+1}} \mathbb{E}_{\bm{\pi}_{\bm{new}}} [\gamma r_{t+2} \\ &+ {\gamma^2} \mathbb{E}_{s_{t+2}}{Q^{\bm{\pi}_{\bm{old}}}(s_{t+2},{\bm{\pi}_{\bm{new}}}(s_{t+2}))}|]\\
&\leq r_{t+1}+\mathbb{E}_{s_{t+1}} \mathbb{E}_{\bm{\pi}_{\bm{new}}} [\gamma r_{t+2} + {\gamma^2}r_{t+3} + ...]\\
&= r_{t+1}+\mathbb{E}_{s_{t+1}} V^{\bm{\pi}_{\bm{new}}}(s_{t+1})\\
&=Q^{\bm{\pi}_{\bm{new}}}
\end{aligned}
\label{policy_improvement}
\end{equation} \hfill $\blacksquare$

Given \textit{Proposition 3} and \textit{Proposition 4}, we can now analyze the convergence of the QuaDRED.

\textit{Theorem 5 (Convergence)}: Let $\bm{\pi}^{\bm{i}}$ be the policy in the $i$-th policy improvement, $i=1,2,...,\infty$, and $\bm{\pi}^{\bm{i}} \rightarrow \pi^{*}$ when $i\rightarrow\infty$. There exists $Q^{\bm{\pi}^{*}}(s, a) \geq Q^{\bm{\pi}^{\bm{i}}}(s, a)$, $\forall s\in \mathcal{S}$ and $\forall a\in \mathcal{A}$.

\textit{Proof}: Since \textit{Proposition 4} suggests $Q^{\bm{\pi}_{\bm{i+1}}}(s, a) \geq Q^{\bm{\pi}_{\bm{i}}}(s, a)$, the sequence $Q^{\bm{\pi}_{\bm{i}}}(s, a)$ is monotonically increasing where $i \in \mathbb{N}$ is a the policy iteration step. Furthermore, \textit{Lemma 2} implies that the the state-action distribution $Z$ over $\mathbb{R}$ has bounded $p$-th moment, so the first moment of $Z$, i.e., $Q^{\bm{\pi}_{\bm{i}}}(s, a)$, is upper bounded. Therefore, the sequence $Q^{\bm{\pi}_{\bm{i}}}(s, a)$ converges to an upper limit $Q^{\bm{\pi}_{*}}(s, a)$ with $\forall s\in \mathcal{S}$ and $\forall a\in \mathcal{A}$. \hfill $\blacksquare$ 

\subsection{Stability Guarantee of the Controller}
In this subsection, the closed-loop stability of QuaDRED-SMPC control framework will be demonstrated. The closed-loop stability is analyzed under the Lipschitz Lyapunov function \cite{rifford2000existence} to guarantee ISS. Before the closed-loop stability analysis, the convexity and Lipschitz continuity of the cost function $\mathcal{L}_{\bm{N}}(\bm{x}_t,\bm{M}_t,\bm{v}_t)$ are introduced in \textit{Proposition 6} and \textit{Proposition 7}, respectively. Since the output of QuaDRED is non-zero-mean and bounded values, which are different from the assumption of zero-mean disturbances in most previous work (\cite{zhang2021stochastic} and \cite{goulart2008input}), the following proofs are all based on the non-zero-mean and bounded disturbances.

We first define an optimal control policy based on the affine disturbance feedback control law:
\begin{equation}
(\bm{M}^{*}(x),\bm{v}^{*}(x)):=\underset{(\bm{M},\bm{v})\in \mathcal{V}_N}{\rm{min}} \mathcal{L}_N(\bm{x},\bm{M},\bm{v})
\label{optimal_control_policy}
\end{equation}

where $\mathcal{V}_N$ is the set of feasible policies, and $(\bm{M}^{*}(x),\bm{v}^{*}(x))$ is a optimal control policy group. The optimal value function $\mathcal{L}^{*}_N(x)$ under the affine disturbance feedback control law is defined as:
\begin{equation}
\mathcal{L}^{*}_N(x):=\underset{(\bm{M},\bm{v})\in \mathcal{V}_N}{\rm{min}} \mathcal{L}_N(\bm{x},\bm{M},\bm{v})
\label{optimal_value}
\end{equation}

Then we demonstrate that the optimal value function $\mathcal{L}_N(x)$ is convex (see \textit{Proposition 6}), so that  \autoref{optimal_value} can be operated as a convex optimization problem.

\textit{Proposition 6}: The function $\mathcal{L}_N(\bm{x},\bm{M},\bm{v})$ is convex.

\textit{Proof}: In \autoref{predicted_cost}, the second term  $\mathbb{E}[\left \|(H_{u}{\bm{M}}\mathcal{G}+H_{w})\bm{w}\right \|^{2}_2]$, i.e., the expected value of a quadratic form with respect to the vector-valued random variable $\bm{w}$, is equal to:
\begin{equation}
\begin{aligned}
&\mathbb{E}[\left \|(H_{u}{\bm{M}}\mathcal{G}+H_{w})\bm{w}\right \|^{2}_2]=\mathbb{E}[{\rm{tr}}((H_{u}{\bm{M}}\mathcal{G}+H_{w})\bm{w} \bm{w}^T)]\\
&={\rm{tr}}((H_{u}{\bm{M}}\mathcal{G}+H_{w})\mathbb{E}[\bm{w} \bm{w}^T])\\
&={\rm{tr}}((H_{u}{\bm{M}}\mathcal{G}+H_{w})({\rm{Cov}}(\bm{w})+\bm{\mu} \bm{\mu}^T))\\
&={\rm{tr}}(\bm{C}^{\frac{1}{2}}_{\bm{w}}(H_{u}{\bm{M}}\mathcal{G}+H_{w})^{T}(H_{u}{\bm{M}}\mathcal{G}+H_{w})\bm{C}^{\frac{1}{2}}_{\bm{w}})\\
&+\bm{\mu}^T(H_{u}{\bm{M}}\mathcal{G}+H_{w})\bm{\mu}
\end{aligned}
\label{convex_simplied_w}
\end{equation}
where $\rm{tr}(\cdot)$ denotes the trace of a square matrix. $\bm{\mu}=\mathbb{E}(\bm{w})$ is the expected value of $\bm{w}$, and $\bm{C}_{\bm{w}}={\rm{Var}}(\bm{w})$ is the variance-covariance matrix of $\bm{w}$. Therefore, $\mathcal{L}_N(x)$ can be written as:
\begin{equation}
\begin{aligned}
&\mathcal{L}_N(x)=\left \|H_{x}x+H_{u}\bm{v}\right \|^{2}_2+\left \|\bm{\mu}\right \|^{2}_{(H_{u}{\bm{M}}\mathcal{G}+H_{w})}\\
&+{\rm{tr}}(\bm{C}^{\frac{1}{2}}_{\bm{w}}(H_{u}{\bm{M}}\mathcal{G}+H_{w})^{T}(H_{u}{\bm{M}}\mathcal{G}+H_{w})\bm{C}^{\frac{1}{2}}_{\bm{w}})
\end{aligned}
\label{convex_l_n}
\end{equation}
where $\left \|x\right \|_{P}$ denotes weighted $2$–norm of the vector $x$. \autoref{convex_l_n} is convex since it consists of convex functions of vector and matrix norms.\hfill $\blacksquare$ 

\textit{Proposition 7}: The function $\mathcal{L}^{*}_N(\bm{x},\bm{M},\bm{v})$ is Lipschitz continuous.

\textit{Proof}: The cost function $\mathcal{L}_N(\bm{x},\bm{M},\bm{v})$ is proved to be convex in \textit{Proposition 6} so that $\mathcal{L}^{*}_N(\bm{x},\bm{M},\bm{v})$ is convex if $\mathcal{V}_N$ has a non-empty interior (\textit{Proposition 1} of \cite{goulart2008input}). $Z$ is a compact (closed and bounded) set so that $\mathcal{L}^{*}_N(\bm{x},\bm{M},\bm{v})$, defined under the compact space $Z$, is piecewise quadratic (\textit{Corollary 4.6} of \cite{goulart2007affine}). Therefore $\mathcal{L}^{*}_N(\bm{x},\bm{M},\bm{v})$ is a Lipschitz continuity function. \hfill $\blacksquare$ 

The above results lead directly to our final result:

\textit{Theorem 9}: Let $\mathcal{W}$, $\mathcal{Z}$ and $\mathcal{X}_{f}$ be polytopes. The closed-loop system (\autoref{linear_discrete_system}) under the SADF control law $\bm{u}_{i|t}$ (in \autoref{affine_fb_SADF}) is ISS. The ISS is also guaranteed in such cases: the stochastic disturbance $\bm{w}_t$ is i.i.d. a bounded and non-zero-mean distribution, i.e., $\mathbb{E}{(w_k)}\neq0$.

\textit{Proof}: According to \textit{Proposition 6} and \textit{Proposition 7}, we first state that the optimal value function $\mathcal{L}^{*}_N(x)$ is a Lipschitz continuity function. The key is then to prove that there exists a Lipschitz continuous function, i.e., $\mathcal{L}^{*}_N(x)$, to satisfy the Lipschitz-ISS criterion (\textit{Proposition 4.15} in \cite{goulart2008input}).

According to \textit{Proposition 1}, there exists a baseline control law $\bm{u}_b$ ensuring ISS under zero-mean distribution disturbance. Let $V_b (x)=V^*_{Nb}(x)-V^*_{Nb}(0)$ be the Lipschitz continuous Lyapunov function \cite{goulart2008input}, where $V^*_{Nb}(x)$ is the optimal value function under the baseline control law $\bm{u}_b$. There exists:
\begin{subequations}
\begin{align}
\alpha_1(\left \| x \right\|) \leq &V_b(x) \leq \alpha_2(\left \| x \right\|) \label{ISS_baseline_control_law_a}\\
V_b(f(x,0)) - &V_b(x) \leq -\alpha_3(\left \| x \right\|) \label{ISS_baseline_control_law_b}
\end{align}
\end{subequations}

Let $V(x)=\mathcal{L}^*_{N}(x)-\mathcal{L}^*_{N}(0)$, where $\mathcal{L}^*_{N}(x)$ is optimal value function under the affine disturbance feedback control law with bounded and non-zero-mean distribution disturbance (in \autoref{optimal_value}). According to \autoref{convex_l_n}, $L^*_{N}(x)$ is shown as:
\begin{equation}
\begin{aligned}
&\mathcal{L}^*_{N}(x)={\rm{min}} \left.\{\mathcal{L}_{N}(x) \right.\}\\
&={\rm{min}}\left.\{ \left \|H_{x}x+H_{u}\bm{v}\right \|^{2}_2 +\left \|\bm{\mu}\right \|^{2}_{(H_{u}{\bm{M}}\mathcal{G}+H_{w})} \right.\\
&\left.+{\rm{tr}}(\bm{C}^{\frac{1}{2}}_{\bm{w}}(H_{u}{\bm{M}}\mathcal{G}+H_{w})^{T}(H_{u}{\bm{M}}\mathcal{G}+H_{w})\bm{C}^{\frac{1}{2}}_{\bm{w}}) \right.\}\\
&=V^*_{Nb}(x)+{\rm{min}}\left.\{\left \|\bm{\mu}\right \|^{2}_{(H_{u}{\bm{M}}\mathcal{G}+H_{w})} \right.\}
\end{aligned}
\label{ISS_opyimal_value}
\end{equation}
where $\bm{\mu}$ is the expected value of disturbances, which is independent with $\bm{v}$ and $\bm{M}$. $V(x)=\mathcal{L}^*_{N}(x)-\mathcal{L}^*_{N}(0)=V^*_{Nb}(x)-V^*_{Nb}(0)=V_b (x)$. Hence, there exists $\mathcal{H}_{\infty}$-functions $\alpha_1(\cdot)$, $\alpha_2(\cdot)$ such that \autoref{ISS_baseline_control_law_a} holds with $V_b(\cdot)=V(\cdot)=\mathcal{L}^*_{N}(\cdot)-\mathcal{L}^*_{N}(0)$.

To prove $V(\cdot)$ satisfying \autoref{ISS_baseline_control_law_b}, note that $V_b(f(x,0)) - V_b(x)=[V^*_{Nb}(f(x,0))-V^*_{Nb}(0)]-[V^*_{Nb}(x)-V^*_{Nb}(0)]=V^*_{Nb}(f(x,0))-V^*_{Nb}(x)$, so that $V^*_{Nb}(f(x,0))-V^*_{Nb}(x) \leq -\alpha_3(\left \| x \right\|)$. It follows that:
\begin{equation}
\begin{aligned}
V&(f(x,0)) - V(x) \\
&= [\mathcal{L}^*_{N}(f(x,0))-\mathcal{L}^*_{N}(0)]-[\mathcal{L}^*_{N}(x)-\mathcal{L}^*_{N}(0)]\\
&=\mathcal{L}^*_{N}(f(x,0))-\mathcal{L}^*_{N}(x)
\end{aligned}
\label{ISS_V_second_inequality}
\end{equation}
where both $\mathcal{L}^*_{N}(f(x,0))$ and $V^*_{Nb}(f(x,0))$ have $w=0$. The only difference between two control laws is zero-mean or non-zero-mean disturbance distributions so that $\mathcal{L}^*_{N}(f(x,0))=V^*_{Nb}(f(x,0))$. Hence, combining with \autoref{ISS_opyimal_value}, \autoref{ISS_V_second_inequality} can be rewritten as:
\begin{equation}
\begin{aligned}
V&(f(x,0)) - V(x) \\
&=V^*_{Nb}(f(x,0))-V^*_{Nb}(x)-{\rm{min}}\left.\{ \left \|\bm{\mu}\right \|^{2}_{(H_{u}{\bm{M}}\mathcal{G}+H_{w})} \right.\}
\end{aligned}
\label{ISS_V_second_inequality1}
\end{equation}

According to \textit{Proposition 6}, we have ${\rm{min}}\left.\{\left \|\bm{\mu}\right \|^{2}_{(H_{u}{\bm{M}}\mathcal{G}+H_{w})}\right.\} \geq 0$. Then we have:
\begin{equation}
\begin{aligned}
V&(f(x,0)) - V(x) \\
&\leq V^*_{Nb}(f(x,0))-V^*_{Nb}(x) \leq -\alpha_3(\left \| x \right\|)
\end{aligned}
\label{ISS_V_second_inequality2}
\end{equation}

\autoref{ISS_V_second_inequality2} above shows that there exists $\mathcal{H}_{\infty}$-functions $\alpha_3(\cdot)$ such that \autoref{ISS_baseline_control_law_b} holds with $V(\cdot)=\mathcal{L}^*_{N}(\cdot)-\mathcal{L}^*_{N}(0)$. Therefore, $V(x)=\mathcal{L}^*_{N}(x)-\mathcal{L}^*_{N}(0)$ is a Lipschitz continuous Lyapunov function, and the ISS of the closed-loop system (\autoref{linear_discrete_system}) is guaranteed with bounded and non-zero-mean distribution disturbances, i.e., $\mathbb{E}{(w_k)}\neq0$. \hfill $\blacksquare$

\section{Numerical Example}
The performance of our proposed QuaDRED-SMPC framework is evaluated in RotorS \cite{furrer2016rotors}, a UAV software simulator. Based on the benchmark \cite{wang2022kinojgm,zhang2019quota}, the parameters of our proposed framework are summarized in \autoref{model_parameters}.

\begin{table}[!t]
    \caption{Parameters of QuaDRED-SMPC}
    \label{model_parameters}
    \begin{tabular}{c c c}
    \toprule
    \textbf{Parameters} & \textbf{Definition} & \textbf{Values} \\
    \hline
    $l_{\mu}$ & Learning rate of actor & 0.001 \\
    $l_{\theta}$ & Learning rate of critic & 0.001 \\
    $\mu$ & \makecell[c]{Actor neural network: fully connected with two \\hidden layers (128 neurons per hidden layer)} & - \\
    $\theta$ & \makecell[c]{Critic neural network: fully connected with two \\hidden layers (128 neurons per hidden layer)} & - \\
    $D$ & Replay memory capacity & $10^6$ \\
    $B$ & Batch size & 256 \\
    $\gamma$ & Discount rate & 0.998 \\
    - & Training episodes & 1000 \\ 
    $T_s$ & MPC Sampling period & 50ms \\
    $N$ & Time steps & 20 \\
    \bottomrule
    \end{tabular}
\end{table}

\subsection{Comparative performance of QuaDRED Training}
\begin{figure}[]
  \centering
  \includegraphics[scale=0.18]{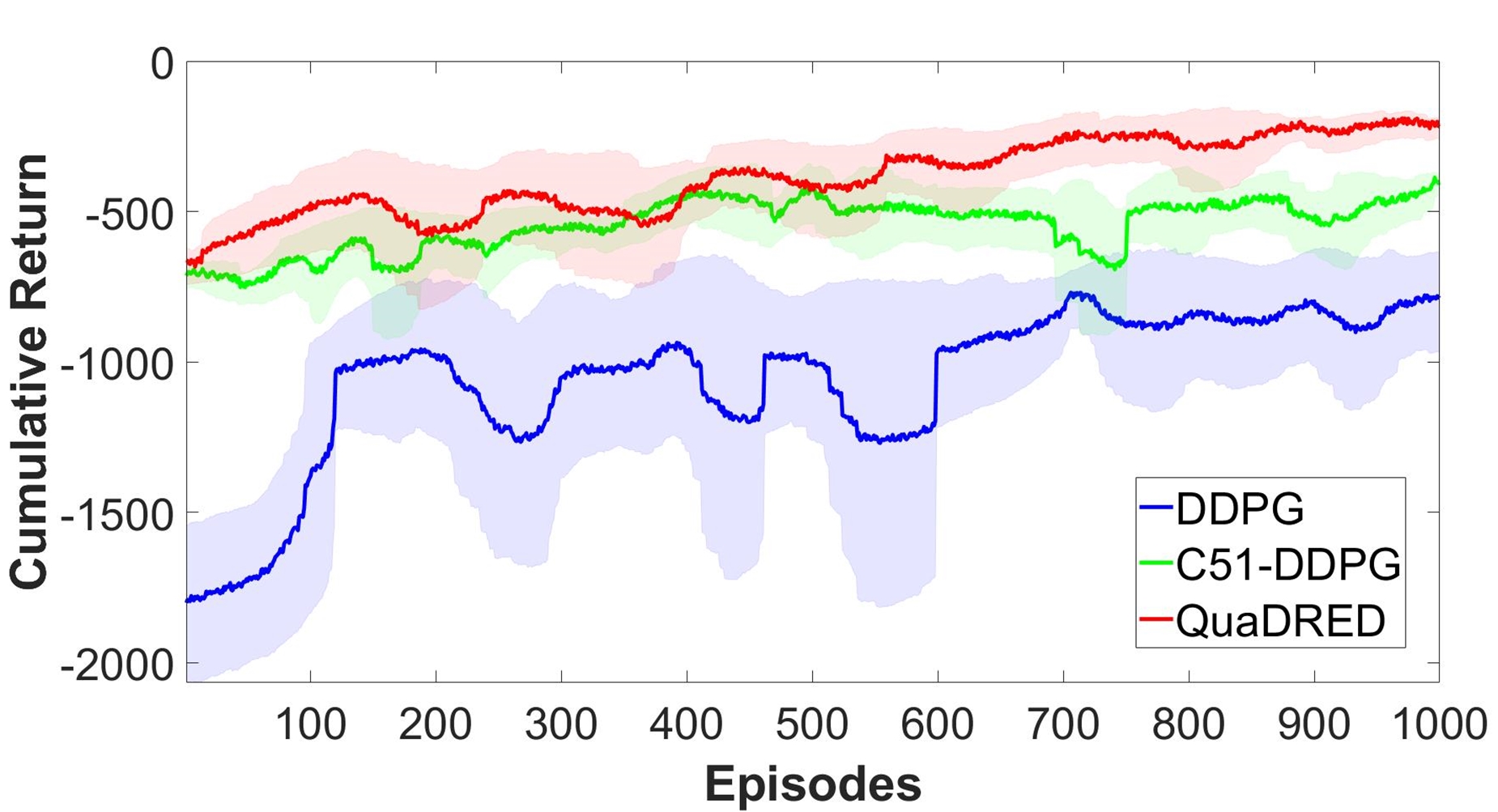}
  \caption{Learning curves of three RL algorithm: DDPG, C51-DDPG and QuaDRED. The simulated speed in Gazebo is set as $0.6$.}
  \label{learning_curves}
\end{figure}

In the training process, the quadrotor system operates with aerodynamic effects in the horizontal plane in the range [-3,3] ($m/s^2$). The quadrotor state $\bm{x}$ is recorded at 16 Hz. The training process is occurs over 1000 iterations. The matrices $H_1$ and $H_2$ in \autoref{reward} are chosen as $H_1=diag\{ 2.5e^{-2}, 2.5e^{-2}, 2.5e^{-2}, 1e^{-3}, 1e^{-3}, 1e^{-3}, 2.5e^{-3}, \\ 2.5e^{-3}, 2.5e^{-3}, 2.5e^{-3}, 1e^{-5}, 1e^{-5}, 1e^{-5}\}$ and $H_2=diag\{ 1.25e^{-4}, 1.25e^{-4}, 1.25e^{-4}, 1.25e^{-4}\}$, respectively. The learning curves are displayed in Fig.~\ref{learning_curves}, where we show the training performance of DDPG \cite{lillicrap2015continuous}, C51-DDPG \cite{bellemare2017distributional} and our proposed QuaDRED. The performance shows that, although all the three algorithms converge to a long-term return eventually, the two distributional RL approaches, C51-DDPG and QuaDRED, outperform the traditional DDPG RL approach. More importantly, in the training process, our proposed QuaDRED achieves the largest return whilst maintaining the highest convergence speed.

\begin{figure}[]
  \centering
  \includegraphics[scale=0.20]{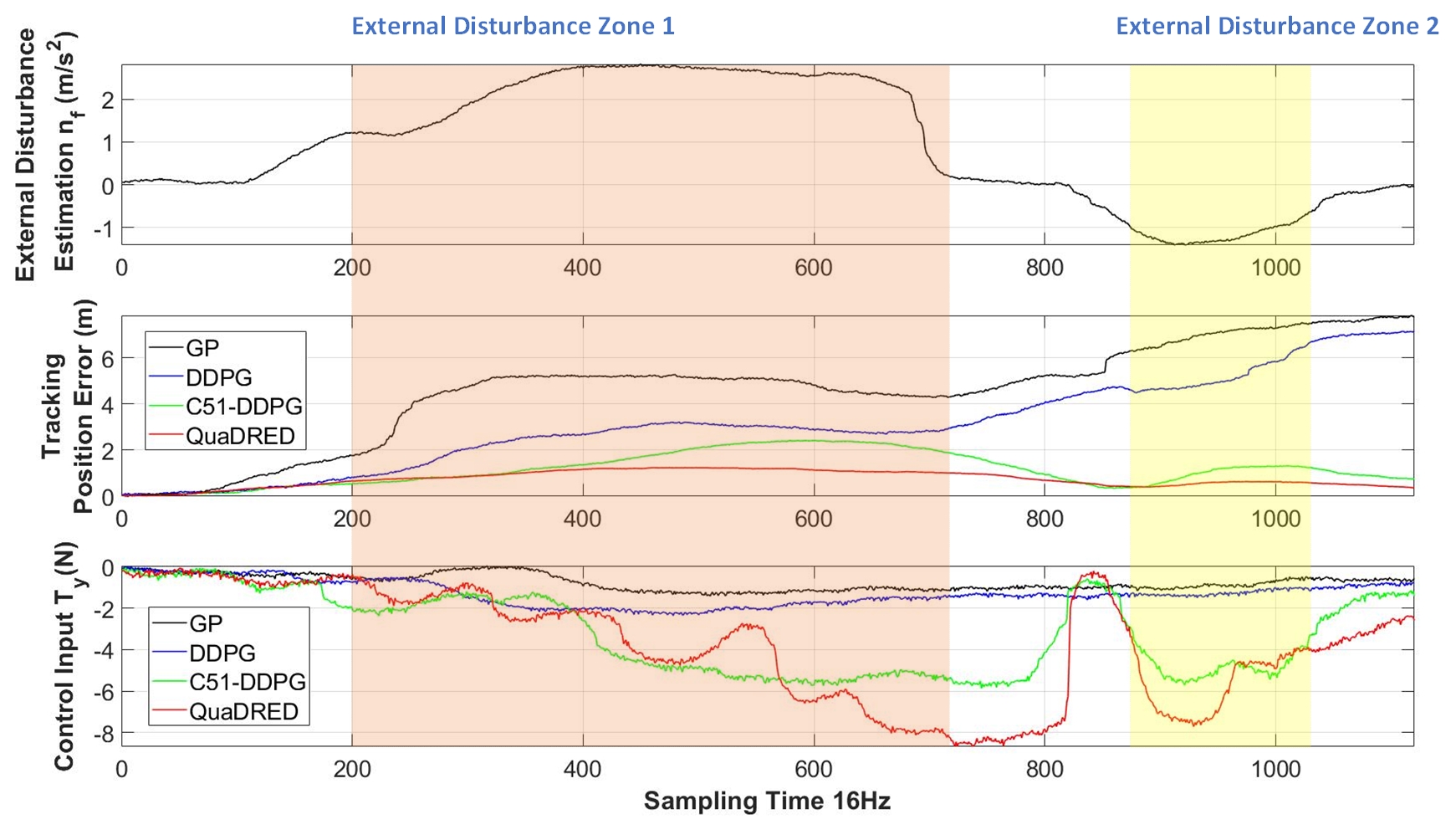}
  \caption{Specific scenarios results: Wind estimation $\bm{e}_{fk}$, position error $(m)$ and Y control input $T_y$ (expressed in body frame).}
  \label{specific_scenarios_data}
\end{figure}

\subsection{Comparative performance of QuaDRED-SMPC under variable aerodynamic effects}
We compare our QuaDRED-SMPC against a state-of-the-art trajectory tracking algorithm, Torrente \cite{torrente2021data}, and interactive approaches, DDPG and C51,  with variable aerodynamic forces added to our simulated environment. The experiments are based on the trained QuaDRED model described in Section \uppercase\expandafter{\romannumeral4}-A. We first set the aerodynamic forces as [0.0, 2.0, 0.0] ($m/s^2$).  Fig.~\ref{specific_scenarios_data} shows the tracking position errors and control input with two opposite heading aerodynamic forces (with the same force [0.0, 2.0, 0.0] ($m/s^2$)), where our proposed QuaDRED-SMPC has the smallest tracking position error, and reacts to the sudden aerodynamic effects sufficiently.

\begin{figure}[]
  \centering
  \includegraphics[scale=0.68]{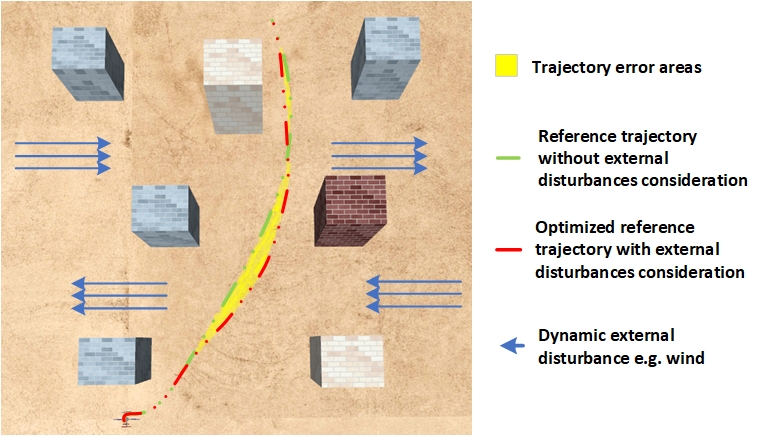}
  \caption{The simulation scenario in RotorS: both reference trajectories with/without external forces are generated by Kino-JSS \cite{wang2022kinojgm}.}
  \label{simulation_env}
\end{figure}

Two larger and more complex forces, i.e., [-2.0, 2.0, 0.0] and [-3.0, 3.0, 0.0] ($m/s^2$), are then used in the scenario shown in Fig.~\ref{simulation_env}. In \autoref{Comparison_of_tracking}, the second approach, `DDPG + MPC', is a combination of DDPG \cite{lillicrap2015continuous} and MPC; and the third approach, `C51 + SMPC', is a combination of C51 \cite{bellemare2017distributional}, DDPG and SMPC. Our results show that interactive approaches are not always better than non-interactive approaches. For example, `DDPG + MPC' has a lower success rate than GP-MPC, whilst there is little difference in the operation time and cumulative tracking error with relatively small aerodynamic forces. However, compared with GP-MPC, our proposed QuaDRED-SMPC achieves improvements of $42\%$, $58\%$ and $71\%$ in operation time, and $86\%$, $75\%$ and $66\%$ in cumulative tracking errors, respectively.

\begin{table}[]
\caption{Comparison of Trajectory Tracking under Programmatic External forces}
\label{Comparison_of_tracking}
\begin{center}
\setlength{\tabcolsep}{1.7mm}{
\begin{tabular}{|l|l|c|c|c|}
\hline
Ex. forces                        & Method      & Succ. Rate & Time (s) & Err. (m) \\ \hline
\multirow{4}{*}{{[}0.0, 2.0, 0.0{]}}  & GP-MPC        & 86\%              & 15.41     & 8.22                        \\ \cline{2-5} 
                                      & DDPG + MPC & 81\%               & 13.57     & 6.89                         \\ \cline{2-5} 
                                      & C51 + SMPC & 90\%             & 9.35   & 1.48                        \\
                                      \cline{2-5}
                                      & QuaDRED-SMPC    & \bf{91\%}             & \bf{9.04}     & \bf{1.19}
                                      \\ \hline
\multirow{4}{*}{{[}-2.0, 2.0, 0.0{]}} & GP-MPC         & 74\%              & 26.11     & 21.80                        \\ \cline{2-5} 
                                      & DDPG + MPC & 78\%               & 18.26     & 15.33                        \\ \cline{2-5} 
                                      & C51 + SMPC & 89\%              & 12.25    & 6.68                        \\
                                      \cline{2-5}
                                      & QuaDRED-SMPC    & \bf{91\%}             & \bf{10.97}     & \bf{5.60}
                                      \\ \hline
\multirow{4}{*}{{[}-3.0, 3.0, 0.0{]}} & GP-MPC         & 17\%              & 41.98     & 35.27                        \\ \cline{2-5} 
                                      & DDPG + MPC & 46\%               & 30.43     & 27.24                        \\ \cline{2-5} 
                                      & C51 + SMPC & 75\%               & 15.31    & 14.53                        \\
                                      \cline{2-5}
                                      & QuaDRED-SMPC    & \bf{82\%}             & \bf{12.22}     & \bf{11.64}
                                      \\ \hline
\end{tabular}
}
\end{center}
\end{table}

\section{Conclusions}
In this paper, we propose an accurate trajectory tracking framework, QuaDRED-SMPC, for quadrotors operating in environments with variable aerodynamic forces. The QuaDRED-SMPC combines aerodynamic disturbance estimation and stochastic optimal control to address the aerodynamic effects on quadrotor tracking. A quantile-approximated distributional RL, QuaDRED, is developed to improve the accuracy of aerodynamic effect estimation, and its convergence is analyzed. Using SADF for control parameterization to guarantee convexity, an SMPC is used to avoid conservative control returns and significantly improves the accuracy of quadrotor tracking. The aerodynamic disturbances are considered to have non-zero mean in the entire QuaDRED-SMPC framework. We demonstrate that our proposed approach can track aggressive trajectories accurately under complex aerodynamic effects whilst guaranteeing both the convergence of QuaDRED and the stability of the whole control framework.
In future works, we will implement QuaDRED-SMPC in real-world flight tests. We will also analyze and evaluate its performance with a greater variety of aerodynamic forces.






\bibliographystyle{unsrt}
\bibliography{ref}

\end{document}